\documentclass[a4paper,fleqn,usenatbib]{mnras}

\usepackage{newtxtext,newtxmath}
\usepackage[T1]{fontenc}
\usepackage{ae,aecompl}

\usepackage{graphicx}	% Including figure files
\usepackage{amsmath}	% Advanced maths commands
\usepackage{mathtools}	% To split equation over two lines
\usepackage{amssymb}	% Extra maths symbols
\usepackage{array}	         % Left justify columns in tables
\usepackage{ragged2e}    % Left justify columns in tables
\usepackage{sidecap}                                         % Allows captions to be printed at the side of a figure instead of underneath it
\sidecaptionvpos{figure}{c}                                  % Vertically centres the side caption 
\usepackage[font=small, labelfont=bf]{caption}   % Makes all the captions small and all the Figure numbers bold
\usepackage{multirow}        				 % Allows for multiline headings and headings that span multiple columns
\usepackage{floatrow}                                         % Should allow me to position the side caption where I want it 
\usepackage{arydshln}					 % Allows for dashed lines to be inserted in tables
\usepackage{bigstrut}					 % Allows lines to be thrown before and after contents of a cell in a table, effectively vertically centering them
\usepackage{float}						% Allows figures to be positioned at exact points in the text
\usepackage[export]{adjustbox}                          % Allows figures to be left/right justified

\graphicspath{ {./Figures/} }               %----- Set default location of figures
\numberwithin{equation}{section}    %----- Numbers equations by section 

\newcommand{\Msun}{{\,M}$_\odot$}
\newcommand{\Lsun}{{\,L}$_\odot$}
\newcommand{\Dsun}{{\,D}$_\odot$}

\newcommand{\kms}{{\,\rm {kms$^{-1}$}}}

\newcommand{\alfab}{{\,[}$\alpha$/H{]}}

\usepackage{longtable}  			% Table covering multiple pages
\usepackage{natbib}                          % To cite the url of a website
\usepackage{csquotes}                      % to put quotation marks around words

\title[Andromeda XXVII and the North West Stream]{A Dwarf Disrupting - Andromeda XXVII and the North West Stream}

\author[J. Preston et al]   
{Janet Preston,$^{1*}$
\ Michelle L.M. Collins,$^1$
\ Rodrigo A. Ibata,$^2$
\ Erik J. Tollerud,$^3$
\newauthor R. Michael Rich$^4$
\ Ana Bonaca,$^5$
\ Alan W. McConnachie$^6$
\ Dougal Mackey$^7$
\newauthor Geraint F. Lewis$^8$
\ Nicolas F. Martin$^{2,9}$
\ Jorge Pe\~{n}arrubia$^{10}$
\ Scott C. Chapman$^{11}$
\newauthor  Maxime Delorme$^{1}$
\\
\\
$^1$Department of Physics, University of Surrey, Guildford, GU2 7XH, Surrey, UK. \thanks{j.preston@surrey.ac.uk} \\
$^2$Observatoire de Strasbourg, 11, rue de l'Universit\'{e}, F-67000, Strasbourg \\
$^3$Space Telescope Science Institute, 3700 San Martin Drive, Baltimore, MD 21218, USA \\
$^4$Department of Physics and Astronomy, University of California at Los Angeles, Los Angeles, CA 90095, USA \\
$^5$Harvard-Smithsonian Center for Astrophysics, 60 Garden St, Cambridge, MA 02138, USA \\
$^6$NRC Herzberg Institute of Astrophysics, 5071 West Saanich Road, Victoria, B.C., V9E 2E7, Canada\\
$^7$Research School of Astronomy and Astrophysics, Australian National University, Canberra, ACT 2611, Australia\\
$^8$Sydney Institute for Astronomy, School of Physics A28, The University of Sydney, NSW, 2006, Australia\\
$^{9}$Max-Planck-Institut f{\"u}r Astronomie, K{\"o}nigstuhl 17, D-69117 Heidelberg, Germany \\
$^{10}$Institute for Astronomy, University of Edinburgh, Royal Observatory, Blackford Hill, Edinburgh EH9 3HJ, UK \\
$^{11}$Department of Physics and Atmospheric Science, Dalhousie University, 6310 Coburg Road, Halifax, B3H 4R2, Canada \\
}

\date{Accepted XXX. Received YYY; in original form ZZZ}

\pubyear{2017}

\begin{document}
\label{firstpage}
\pagerange{\pageref{firstpage}--\pageref{lastpage}}
\maketitle

\begin{abstract}
We present a kinematic and spectroscopic analysis of 38 red giant branch stars, in 7 fields, spanning the dwarf spheroidal galaxy Andromeda XXVII and the upper segment of the North West Stream.  Both features are located in the outer halo of the Andromeda galaxy at a projected radius of 50-80 kpc, with the stream extending for $\sim$3$^{\circ}$ on the sky. Our data is obtained as part of the PAndAS survey and enables us to confirm that Andromeda XXVII's heliocentric distance is 827 $\pm$ 47 kpc and spectroscopic metallicity is  -2.1$^{+0.4}_{-0.5}$. We also re-derive Andromeda XXVII's kinematic properties, measuring a systemic velocity = -526.1$^{+10.0}_{-11.0}${\kms} and a velocity dispersion that we find to be non-Gaussian but for which we derive a formal value of 27.0$^{+2.2}_{-3.9}${\kms}. In the upper segment of the North West Stream we measure mean values for the metallicity = -1.8$\pm$0.4, systemic velocity = -519.4 $\pm$4.0{\kms} and velocity dispersion = 10.0$\pm$4.0{\kms}. We also detect a velocity gradient of 1.7$\pm$0.3 {\kms} kpc$^{-1}$ on an infall trajectory towards M31. With a similar gradient, acting in the same direction, in the lower segment we suggest that the North West Stream is not a single structure. As the properties of the upper segment of the North West Stream and Andromeda XXVII are consistent within ~90\% confidence limits, it is likely that the two are related and plausible that Andromeda XXVII is the progenitor of this stream.

\end{abstract} 

\begin{keywords}
galaxies: dwarf -- galaxies: fundamental parameters -- galaxies: kinematics and dynamics -- Local Group
\end{keywords}

%%%%%%%%%%%%%%%%%%%%%%%%%%%%%%%%%%%%%%%%%%%%%%%%%%
%%%%%%%%%%%%%%%%% BODY OF PAPER %%%%%%%%%%%%%%%%%%
%--------------------------------------------------------------------------- Introduction -------------------------------------------------------------------------------------------------
\section{Introduction} \label{introduction}
\graphicspath{ {Figures/} }    %----- Set default location of figures 

% Figure placed here to appear in the right place in the pdf
\begin{figure*}
	\includegraphics[height=.5\paperheight, width=.9\paperwidth]{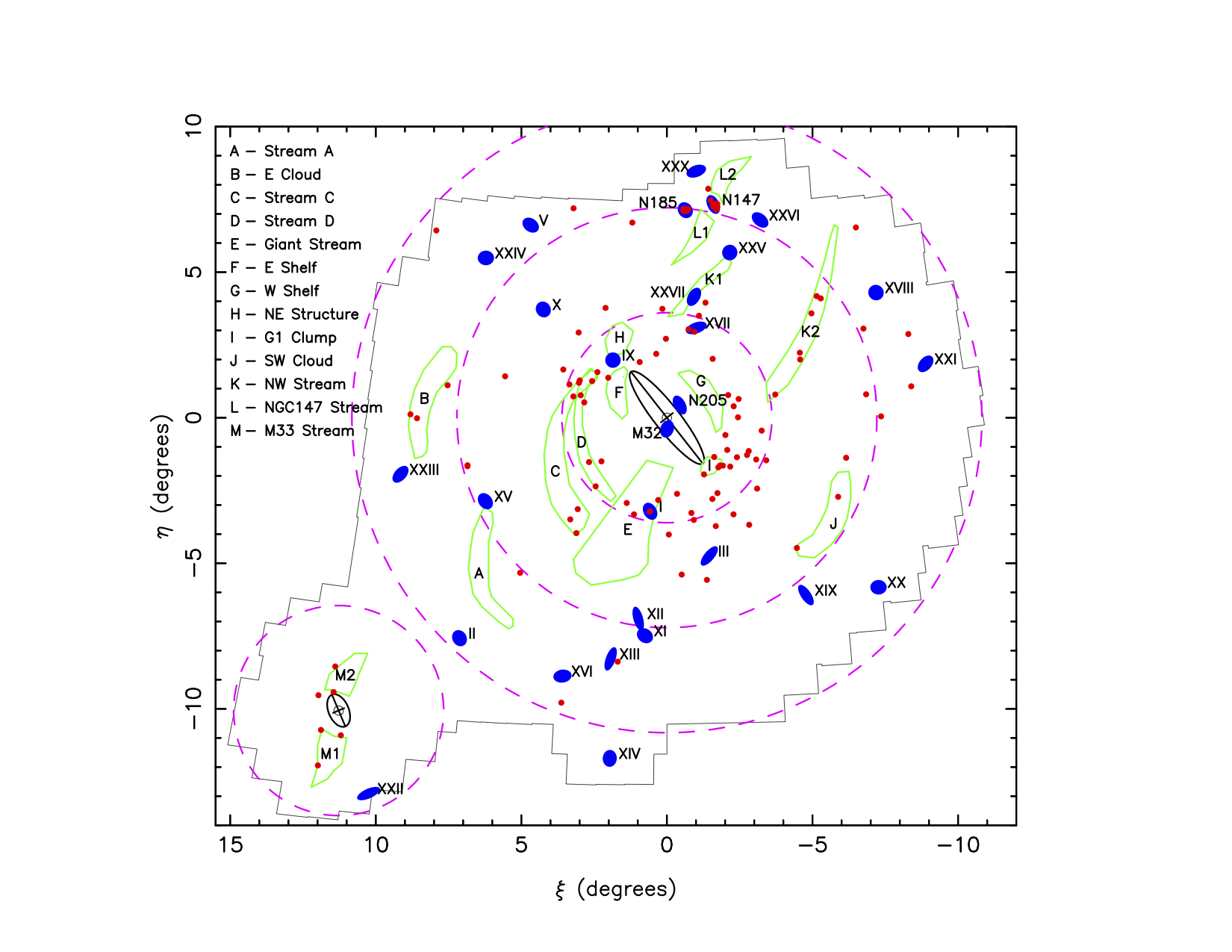}
    	\vspace*{-11mm}\caption[And XXVII and NW stream within the PAndAS footprint]
	{And XXVII and NW stream within the PAndAS footprint. The dashed circles show projected radii of 50kpc, 100kpc and 150kpc from M31, and 50kpc from M33.  The grey polygon denotes the outline of the PAndAS footprint. Red dots indicate known globular clusters at projected radii of greater than 1 degree from M31. Blue ellipses correspond to known dwarf galaxies in the PAndAS footprint. Other stellar substructures in the halo of M31 are outlined in green. Figure reproduced from \cite{RefWorks:449}.}
	\label{Fig1}
\end{figure*}

The spectacular stellar streams snaking around the Milky Way (MW) and Andromeda (M31) are the paleontological remnants of mergers on a galactic scale.  In the past decade, wide and deep photometric surveys, such as the Dark Energy Survey (DES,  \citealt{RefWorks:420}, \citealt{RefWorks:411}), the Pan-Andromeda Archaeological Survey (PAndAS, \citealt{RefWorks:58}), the Sloan Digital Sky Survey (SDSS, \citealt{RefWorks:419}), the Pan-STARRS1 3$\pi$ Survey (PS1, \citealt{RefWorks:562}) and the ESA/Gaia survey (\citealt{RefWorks:204}) have discovered more than 60 streams around the MW (\citealt{RefWorks:406}, \citealt{RefWorks:411},  \citealt{RefWorks:468} and \citealt{RefWorks:545}) and more than 10 streams around M31 (\citealt{RefWorks:47}, \citealt{RefWorks:82}, \citealt{RefWorks:107}, \citealt{RefWorks:449}). 

These streams comprise debris from the tidal disruption of smaller stellar structures, such as dwarf galaxies and globular clusters, as they orbit around their host galaxies. They provide visible evidence that large galaxies grow by assimilating smaller ones.  Using data from stellar streams we can create models to show how small galaxies are accreted by larger ones, the timescales over which this happens and test the $\Lambda$CDM paradigm of hierarchical galaxy formation (\citealt{RefWorks:513}, \citealt{RefWorks:514}, \citealt{RefWorks:515}).  We also know that the stellar debris within a stream follows the orbital path of its progenitor. So, using Newton's law of attraction, we can constrain the gravitational potentials, masses and dark matter distributions of the host galaxies (\citealt{RefWorks:423, RefWorks:176}, \citealt{RefWorks:10},  \citealt{RefWorks:421}, \citealt{RefWorks:327}, \citealt{RefWorks:245}, \citealt{RefWorks:422},  \citealt{RefWorks:82}, \citealt{RefWorks:424}).

In the MW, several streams are presumed to have dwarf galaxy progenitors, for example: the Sagittarius (\citealt{RefWorks:499}, \citealt{RefWorks:363}, \citealt{RefWorks:121}, \citealt{RefWorks:406}), Orphan (\citealt{RefWorks:505}, \citealt{RefWorks:501}, \citealt{RefWorks:534},  \citealt{RefWorks:537}), Cetus Polar (\citealt{RefWorks:504}), PAndAS MW (\citealt{RefWorks:47}) and Styx  (\citealt{RefWorks:503}, \citealt{RefWorks:475}) streams. 

In M31, stream fields are found near NGC147 (\citealt{RefWorks:18}, \citealt{RefWorks:75}) and NGC205 (\citealt{RefWorks:190}, \citealt{RefWorks:107}).   In the inner halo, the Giant Stellar Stream, (\citealt{RefWorks:143}, \citealt{RefWorks:148}, \citealt{RefWorks:176}, \citealt{RefWorks:483}, \citealt{RefWorks:11}) the North-Eastern Shelf, labelled the NE Structure in Figure \ref{Fig1}, (\citealt{RefWorks:500}, \citealt{RefWorks:60}) and the Western Shelf (\citealt{RefWorks:147, RefWorks:446, RefWorks:59} and \citealt{RefWorks:155}) are thought to be the tidal debris, from multiple pericentric passages of the M31 centre, of a progenitor with an estimated stellar mass $\sim$1-5 x 10$^9$M$_{\sun}$. In the outer halo, questions still surround the progenitor of the luminous South West Cloud  (\citealt{RefWorks:12}, \citealt{RefWorks:107}) while the progenitor of the North West (NW) Stream is thought to be Andromeda XXVII (And XXVII), \cite{RefWorks:18}, \cite{RefWorks:56},  \cite{RefWorks:42},  \cite{RefWorks:206}.

The NW Stream comprises two segments. The lower segment (labelled K2 in Figure \ref{Fig1} and hereafter referred to as NW-K2) was discovered by \cite{RefWorks:58} using PAndAS data obtained from the 3.6 m Canada-France-Hawaii Telescope (CFHT). They find it to be $\sim$6$^{\circ}$ ($\sim$ 80 kpc) long in projection at a projected radius of $\sim$50-120 kpc from the centre of M31.  The upper segment (labelled K1 in Figure \ref{Fig1} and hereafter referred to as NW-K1) was discovered two years later by \cite{RefWorks:18}. They report this segment to be almost $\sim$3$^{\circ}$ long in projection at a projected radius $\sim$50-80 kpc from the centre of M31.  Despite the two segments being quite separate, see Figure \ref{Fig1}, based on their morphology \cite{RefWorks:18} consider them to be part of a single stellar structure entwined around M31. This view is supported by  \cite{RefWorks:82}, who detected similar metallicities in both segments of the stream, and by \cite{RefWorks:76}. Their work indicates that the stars in the stream are $\sim$10 Gyrs old, that the stream is $\sim$5 kpc wide and that it extends for a projected distance $\sim$200 kpc, making it one of the longest in the Local Group (c.f. the MW's Sagittarius Stream, which is detected at heliocentric distances of 37 kpc $\le$ {\Dsun} $\le$ 117 kpc, implying an estimated length $\sim$ 80 kpc, \cite{RefWorks:600}, and the Giant Stellar Stream, to the South East of M31, with an estimated length of $\sim$ 100 kpc, \cite{RefWorks:11}).  \cite{RefWorks:76} report that NW-K2 is almost complete while NW-K1 contains a number of clearly visible gaps that could have been induced by dark matter sub-halos. Recent modelling of NW-K2 by \cite{RefWorks:313} indicates that a progenitor would need to have a stellar mass $\sim$10$^{6-8}${\Msun} and a minimum r$_h$ $\ge$ 30 pc, which is consistent with it being a dwarf galaxy, and that the dSph And XXVII could be the progenitor of the full stream.

And XXVII was discovered, contemporaneously with NW-K1, by \cite{RefWorks:18}.  Kinematic analysis by \cite{RefWorks:42} led these authors to agree with \citeauthor{RefWorks:18} that And XXVII is not in dynamical equilibrium and is no longer a bound system, a view also supported by \cite{RefWorks:206} and \cite{RefWorks:474}.

\begin{table}
	\centering
	\setlength\extrarowheight{2pt}
	\caption[Properties of And XXVII]
	{Properties of AndXXVII as determined by (a) \cite{RefWorks:18} and (b) \cite{RefWorks:42} }		
	\label{table:01}
	\begin{tabular}{ll} 
		\hline 
		Right Ascension(J2000)$^{(a)}$             & 00$^h$:37$^m$:27$^s$.1 $\pm$ 0$^s$.5  \\  [0.5ex]     
		Declination(J2000)$^{(a)}$                     & +45$^o$: 23$^m$ 13$^s$.0 $\pm$ 10$^s$\\[0.5ex] 
		Distance from M31$^{(a)}$                     &  86 $\pm$ 48 kpc\\[0.5ex] 
		M$_v$ $^{(a)}$                                        & -7.9 $\pm$ 0.5 \\[0.5ex] 
		r$_h$$^{(a)}$                                           & 455 $\pm$ 80 pc\\[0.5ex] 
		Fe/H$^{(b)}$                                            & -2.1$\pm$ 0.5  \\[0.5ex] 
		M$_{r_h}$ $^{(b)}$                                  &  8.3$^{+2.8}_{-3.9}$ x 10$^7${\Msun}\\[0.5ex] 
		M/L(r$<r_h$) $^{(b)}$                              & 1391$^{+1039}_{-1128}${\Msun}/{\Lsun}\\[0.7ex] 
		Systemic velocity v$_r$$^{(b)}$              & -539.6$^{+4.7}_{-4.5}$ {\kms} \\[0.5ex] 
		Velocity dispersion $\sigma_v$ $^{(b)}$  & 14.8$^{+4.3}_{-3.1}${\kms} \\[0.5ex] 
		\hline
	\end{tabular}
\end{table}

To provide more insight into this complex and intriguing dSph galaxy and, potentially, its tidal stream, we present in this paper a study of the stellar populations of And XXVII and NW-K1 to determine if the two are associated. Using dynamical data for stars in these features we determine their systemic velocities, velocity dispersions and metallicities. We also show that And XXVII could be the progenitor of NW-K1 and that NW-K1 may be a separate feature from NW-K2.

The paper is structured as follows: Section \ref{Observations} describes our observations and the data reduction process, Section \ref{Analysing membership of And XXVII and NW-K1} describes our data analysis approach and we present our conclusions in Section \ref{Conclusions}. 

% Table placed here to appear in the right place in the pdf
\begin{table*}
	\centering
	\setlength\extrarowheight{2pt}
	\caption[Properties for each observing mask]
	{Properties for each observing mask, including: mask name; date observations were made; observing PI; Right Ascension and Declination of the centre of each mask; projected distance of the centre of the mask from And XXVII ($D_{\rm A27}$) and the number of stars likely to belong to each of the stellar populations (i.e. And XXVII, M31 and the MW) based on probability of membership. The $\alpha$ and $\delta$ for the centre of each mask are determined by taking the mean of the coordinates for all stars on the mask.  The masks are listed in order of increasing distance from M31.}		
	\label{table:1}
	\begin{tabular}{lclccccccc} 
		\hline
		\multirow{2}{*} {Mask name} & 
		\multirow{2}{*} {Date} &
		\multirow{2}{*} {PI} &
		\multirow{2}{*} {$\alpha_{\rm J2000}$} & 
		\multirow{2}{*} {$\delta_{\rm J2000}$}  & 
		\multicolumn{1}{c}{$D_{\rm A27}$}  &     
		\multicolumn{3}{c}{No. of candidate stars within...}\\ [0.5ex] 
		\cline{7-9}
		&  &  &  &  &  kpc  & And XXVII & M31 & MW\\ [0.5ex] 
		\hline 
		A27sf1   &  2015-09-12  &  Collins  &  00:39:39.96  & +45:08:47.73  & 6.6   &   8  &  5  &  57 \\ [0.5ex] 
		603HaS &  2010-09-09  &  Rich      &  00:38:58.52  & +45:17:32.20  & 4.1   &   8  &  4  &  53 \\ [0.5ex] 
		7And27  &  2011-09-26  &  Rich      &  00:37:29.40  & +45:24:12.50  & 0.3   &  11  &  4  &  54 \\ [0.5ex] 
		A27sf2   &  2015-09-12  &  Collins  &  00:36:13.17  & +45:32:31.68  & 3.8   &    2  &  8  &  49 \\ [0.5ex]  
		A27sf4   &  2015-09-12  &  Collins  &  00:33:28.25  & +45:49:24.87  & 11.7  &   0  &  4  &  63 \\ [0.5ex]  
		604HaS &  2010-09-09  &  Rich      &  00:32:05.16  & +46:08:31.20  & 17.4  &   1  &  3  &  68 \\ [0.5ex] 
		A27sf3   &  2015-09-12  &  Collins  &  00:30:25.60  & +46:14:52.66  & 21.6  &   4  &  1  &  76 \\ [0.5ex] 
		\hline
	\end{tabular}
\end{table*}

% Figure placed here to appear in the right place in the pdf
\begin{figure*}
  	\centering
	\includegraphics[height=.4\paperheight, width=.86\paperwidth]{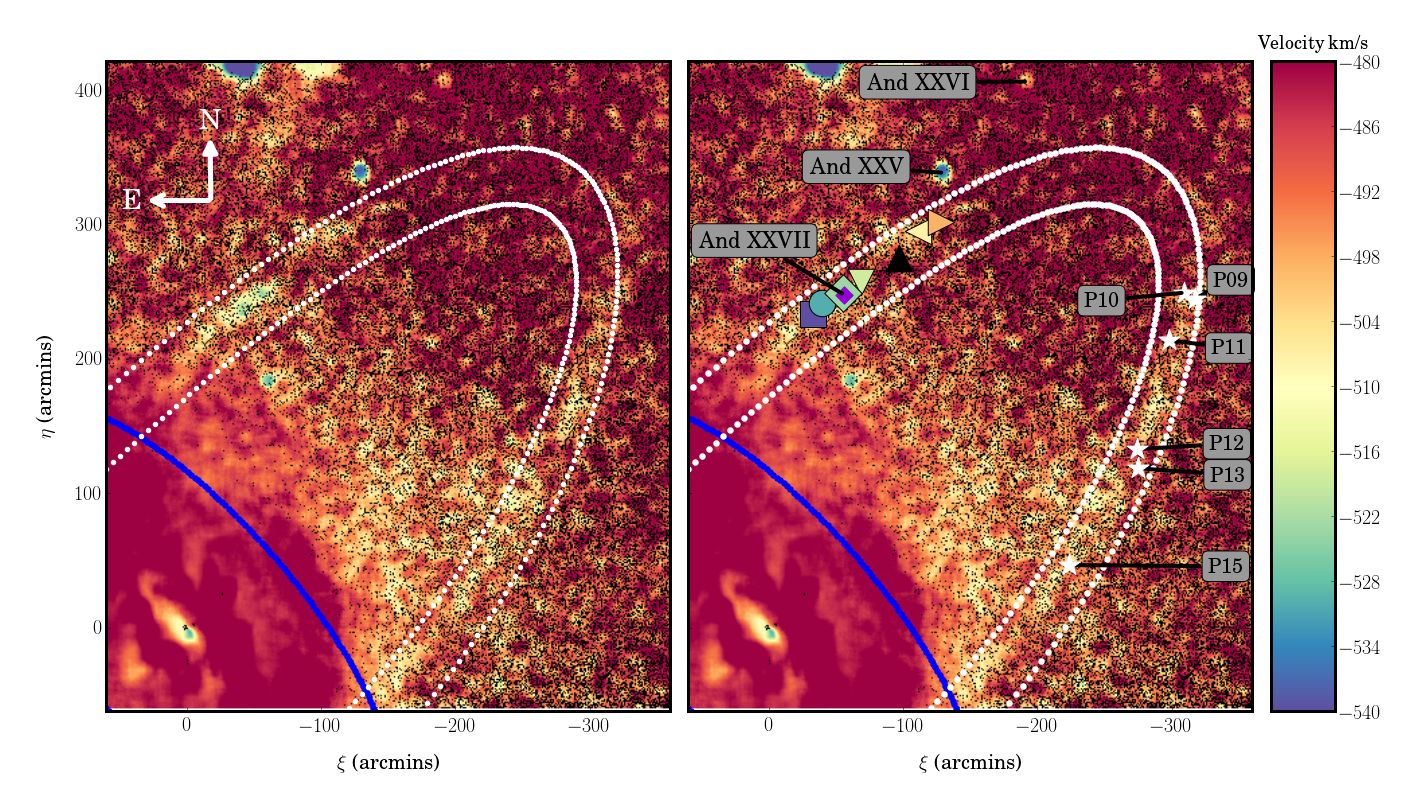}
    	\vspace*{-6mm}\caption[Approximate on-sky positions of the masks, And XXVII and the NW Stream]
	{Density plot of RGB stars in the area around And XXVII and the NW stream.  The data are obtained from the PAndAS catalogue for stars with 20.5 $\le$ \textit{i$_0$} $\le$ 24.5 and -2.0 $\le$ [Fe/H] $\le$ -0.5 and plotted in tangent plane standard coordinates centred on M31. The solid blue line represents the M31 halo (taking a semi-major axis of  55 kpc with a flattening of 0.6, \citealt{RefWorks:107}). The left-hand panel includes dotted, white, lines outlining the inner and outer edges of an ellipse tracing the possible track of the NW Stream, assuming it to be a single feature (following the approach by \citealt{RefWorks:76}). The right-hand panel shows the same data overlaid with the on-sky positions of the observing masks, which are represented by: square = 603HaS; circle = A27sf1; diamond = 7And 27 (inset with a smaller, purple, diamond indicating the centre of And XXVII); inverted triangle = A27sf2; black triangle = A27sf4; left-pointing triangle = 604HaS and right-pointing triangle = A27sf3.  The icons are coloured coded by the systemic velocities derived later in this work, except for A27sf4, for which no And XXVII/NW-K1 candidates were found and so is coloured black.  The Figure also indicates the relative positions of two other M31 satellites, And XXV and And XXVI (\citealt{RefWorks:18}) and globular clusters, PAndAS-09 (P09) - PAndAS-15 (P15) in NW-K2  (\citealt{RefWorks:236}).
	}
	\label{Fig2}
\end{figure*}

%%%%%%%%%%%%%%%%%%%%%%%%%%%%%%%%%%%%%%%%%%%%%%%%%%%%%%%%%%%%%%%%%%%%%%
%%%%%%%%%%%%%%%%%%%%%%%%%%%%%%%%%%%%%%%%%%%%%%%%%%%%%%%%%%%%%%%%%%%%%%
%-----------------------------------------------------------Section 2 - Observations -------------------------------------------------------------------------------------------------
\section{Observations} \label{Observations}
\graphicspath{ {Figures/} } 

%--------------------------------------------------------------------------------
\subsection{Photometry - CFHT}
Data for the initial observations were obtained as part of the PAndAS survey.  PAndAS used the 3.6 m CFHT with the MegaPrime/MegaCam camera, comprising 36, 2048 x 4612, CCDs with a pixel scale of 0.185"/pixel, delivering an almost 1 degree$^2$ field of view (\citealt{RefWorks:58}). To enable good colour discrimination of Red Giant Branch (RGB) stars, g-band (4140\AA{}- 5600\AA) and i-band (7020\AA{}-8530\AA) filters were used.  Good seeing of < 0".8 enabled individual stars to be resolved to depths of \textit{g} = 26.5 and \textit{i} = 25.5 with a signal to noise ratio of $\sim$10 (\citealt{RefWorks:58}, \citealt{RefWorks:42}, \citealt{RefWorks:81}).   

The data were initially processed by the Elixir system, \cite{RefWorks:570}, at CFHT, which, in addition to ascertaining the photometric zero points, also de-biased, flat-fielded and fringe-corrected it.  The data were then transferred to the Cambridge Astronomical Survey Unit to be further reduced using the bespoke pipeline described by \cite{RefWorks:336}. Following this, the data were classified morphologically as, e.g., point source, non-point source and noise-like, then stored with band-merged \textit{g} and \textit{i} data (see \citealt{RefWorks:18}). For this work we select point source objects. 

%--------------------------------------------------------------------------------
\subsection{Spectroscopic Observation - Keck DEIMOS}

Data are obtained from seven masks (the on-sky positions of which are indicated in Figure \ref{Fig2}, and the data from which are provided in the on-line version of this paper at Appendix A) spanning a distance of $\sim$120 arcmins ($\sim$30 kpc) of NW-K1 and crossing the centre of And XXVII.  Three of the masks were observed in 2010 and 2011 (603HaS, 604HaS and 7And27) and may have been analysed by \cite{RefWorks:42} and \cite{RefWorks:206}, though they are not explicitly named in either work.  The remaining four masks, A27sf1, A27sf2, A27sf3 and A27sf4, provide us with additional spectroscopic data obtained using the DEep-Imaging Multi-Object Spectrograph (DEIMOS) on the Keck II Telescope on the dates shown in Table \ref{table:1}.  Our observations use the OG550 filter with the 1200 lines/mm grating with a resolution of $\sim$1.1\AA {}-1.6\AA {} at FWHM. Each mask/site is observed for 1 hour, split into 3 x 20 minute integrations. Observations focus on the Calcium Triplet (CaT) region located between wavelengths 8400\AA{} and 8700\AA. 

We prioritise our targets by selecting stars based on their location within the colour magnitude diagram (CMD). Our highest priority targets are bright stars which lie directly on the And XXVII RGB with 20.3 < \textit{i$_0$} < 22.5. Our next priority are fainter stars on the RGB, i.e. 22.5 < \textit{i$_0$} < 23.5. We then fill the remainder of the mask with stars with 20.5 < \textit{i$_0$} < 23.5 and 0.0 < \textit{g-i} < 4.0.

The data are reduced using a specifically constructed pipeline, described in \cite{RefWorks:216}, that corrects for: scattered light, flat-fields, the slit function and illumination within the telescope and calibrates the wavelength of each pixel. The final phase of the pipeline determines the velocities and associated uncertainties for the stars by: \textit{(1)} creating model spectra comprising a continuum and the absorption profiles of the CaT lines (at 8498\AA, 8542\AA{} and 8662\AA); \textit{(2)} cross-correlating these models with non-resampled stellar spectra using a Markov Chain Monte Carlo (MCMC) approach to obtain the optimum Doppler shift and CaT line widths; and \textit{(3)} correcting the velocities and associated uncertainties, obtained from the posteriors of the above analysis, to the heliocentric frame.

%%%%%%%%%%%%%%%%%%%%%%%%%%%%%%%%%%%%%%%%%%%%%%%%%%%%%%%%%%%%%%%%%%%%%%
%%%%%%%%%%%%%%%%%%%%%%%%%%%%%%%%%%%%%%%%%%%%%%%%%%%%%%%%%%%%%%%%%%%%%%
%---------------------------Section 3:  Determining membership of And XXVII and the NW Stream ----------------------------------------------------------------------------------------
\section{Analysis of And XXVII and NW-K1} \label{Analysing membership of And XXVII and NW-K1}
\graphicspath{ {Figures/} } 

To enable us to determine if And XXVII and NW-K1 are associated we identify and confirm members of their stellar populations, determine the systemic velocities and velocity dispersions of these populations and obtain their metallicities.

\begin{figure*}
	\begin{center}	
	\vspace{-5.0mm}                        % This adjusts the amount of vertical space between the top row of the sub-plots and the top of the page
	\includegraphics[height=.75\paperheight, width=.8\paperwidth]{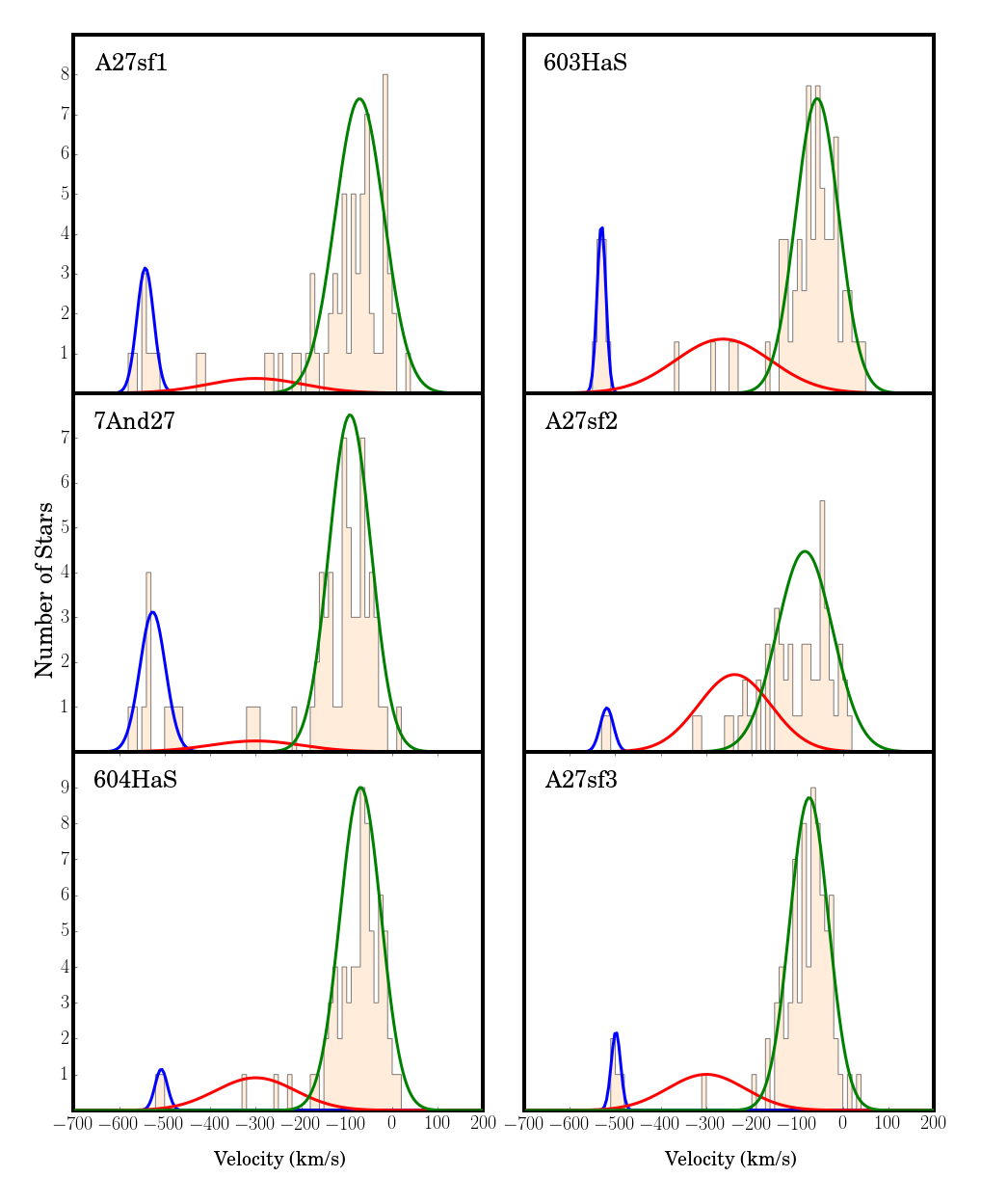}
	\vspace*{-7mm}\caption[Velocity histograms]
	{Kinematic analysis of And XXVII and NW-K1 showing velocity histograms overlaid with the membership pdf for each of the three stellar populations - coloured blue for And XXVII/NW-K1, red for M31 and green for the MW.}
	\label{Fig5}
	\end{center}
\end{figure*}

% Figure placed here to appear in the right place in the text
\begin{figure*}
	\begin{center}
		\vspace{-5.0mm}                        % This adjusts the amount of vertical space between the top row of the sub-plots and the top of the page
		\includegraphics[height=.78\paperheight, width=.8\paperwidth]{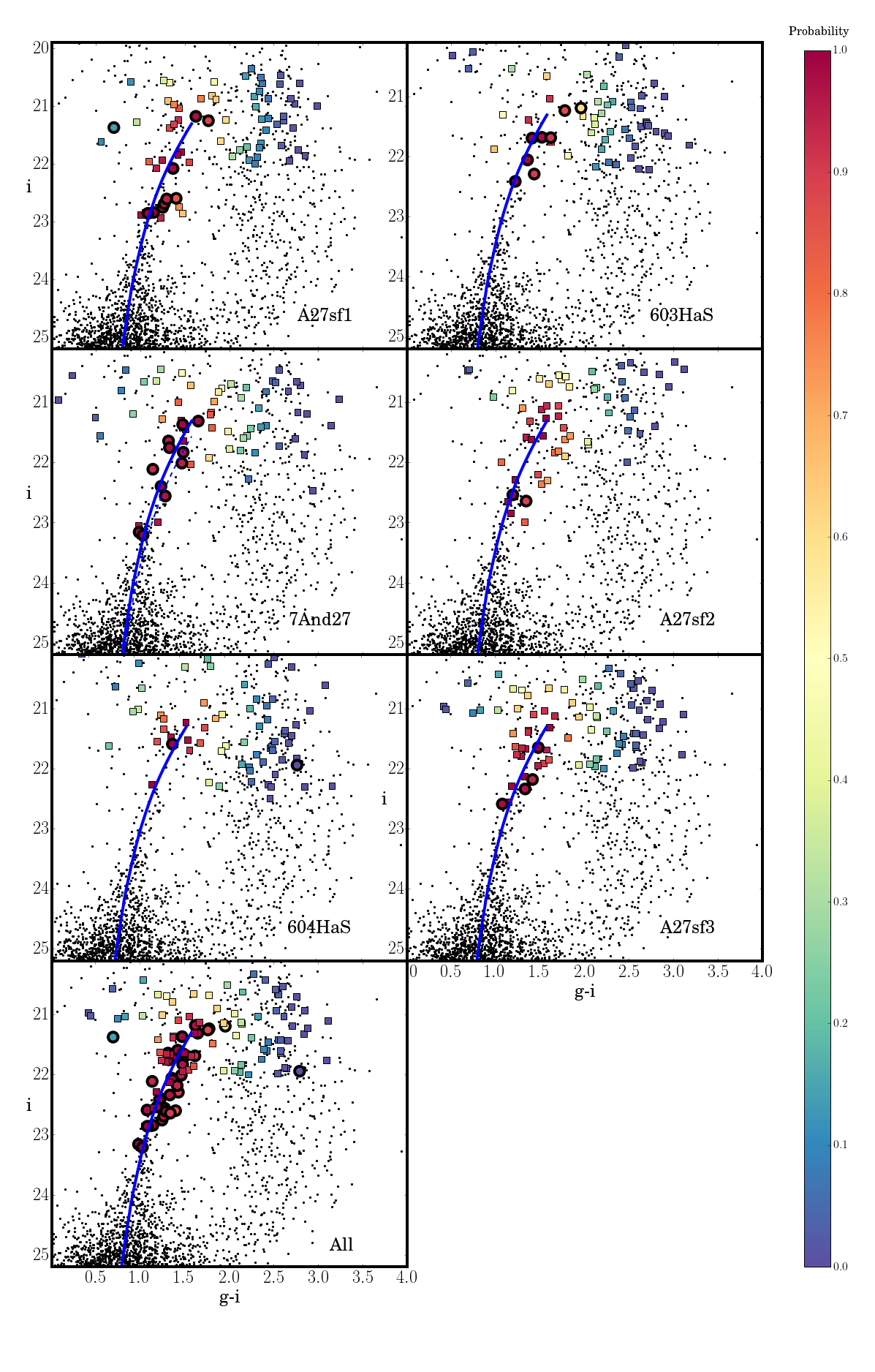}
		\vspace*{-16mm}\caption[CMD for And XXVII masks with 12 Gyr isochrones]
	{CMD for And XXVII/NW-K1 masks with a 12 Gyr, [Fe/H] = -1.7, {\alfab} = 0.0 isochrone, corrected for extinction and an heliocentric distance of 827 kpc. The small black dots show stars from the PAndAS catalogue that lie within 5 arcmins of the centre of the respective masks. The dashed line in the plot for 7And27 shows the position of the same isochrone distance corrected to 1255 kpc (a previously reported heliocentric distance for And XXVII). Round icons represent candidate And XXVII/NW-K1 stars. Square icons represent candidate MW and M31 halo stars. These first cut, candidate populations are based on stellar velocities. The icons are colour coded by probability of membership of And XXVII and NW-K1 based on proximity to the isochrone. The bottom right panel shows the combined results for all masks.}
	\label{Fig3}
	\end{center}
\end{figure*}

%--------------------------------------------------------------------------------
\subsection{Stellar Populations} \label{Stellar Populations}

To avoid any obvious failures of the pipeline and to remove velocities with high uncertainties, we select stars that have velocities in the range -650 {\kms} to 50 {\kms} with velocity uncertainties < 20 {\kms}. Applying these criteria we find no And XXVII/NW-K1 candidate stars on mask A27sf4, which leaves us with a potential gap in the stream. Since this location is targeted on the same basis as the other masks, we conclude that the stream, at this point, must have a lower density than the surrounding areas.  This is consistent with \cite{RefWorks:76} who found significant density variations along the NW-K1 stream.

We plot velocity histograms for the remaining masks, see Figure \ref{Fig5}. These reveal three kinematically distinct stellar populations: stars likely to be members of the MW (v$_r$ =  $\sim$ -80 {\kms}, \citealt{RefWorks:42}), stars likely to be members of the M31 halo (systemic velocity $\sim$ -300 {\kms}, \citealt{RefWorks:57}) and stars consistent with previously published values for the systemic velocity of And XXVII, v$_r$, = -539.6$^{+4.7}_{-4.5}${\kms} (\citealt{RefWorks:42}).

To confirm if the stars in this latter category are members of the And XXVII/NW-K1 population we first look at their proximity to a fiducial isochrone. \cite{RefWorks:18} find the age of And XXVII to be 12 Gyr and its metallicity [Fe/H] $\sim$-1.7, so we select an isochrone with these properties and  {\alfab} = 0.0 from the Dartmouth Stellar Evolution database, \cite{RefWorks:141}. We correct for extinction and the heliocentric distance of And XXVII (827 kpc) and overlay it on the And XXVII RGB. We use the isochrone solely as a fiducial ridge-line for the And XXVII RGB. It is not used to derive or affect any properties reported in this paper, other than a given star's proximity to the RGB. Potential members of And XXVII and NW-K1 should fall along this isochrone, with stars lying close to it more likely to be members than those further away. Following the technique described by \cite{RefWorks:92}, we assign a probability of membership to each star based on its proximity to the isochrone using:
\begin{equation} 
	\label{eq:1}
		P_{\rm iso}  = \rm exp\bigg(\frac{- \Delta (g-i)^2} {2 \sigma_c} - \frac{ \Delta (i)^2} {2 \sigma_m} \bigg)
\end{equation}
where $\Delta(g-i)$ and $\Delta(i)$ are distances from the isochrone and $\sigma$$_c$ (which takes into account the range of colours of the stars on the CMD) and $\sigma$$_m$ (which factors in distance and photometric errors) are free parameters.  \citeauthor{RefWorks:92}'s values were adopted as the starting point and adjusted until values were obtained where stars that lie far from the And XXVII RGB have a low probability of association with the isochrone. We find the optimum values to be: $\sigma$$_c$ = 0.15 and $\sigma$$_m$ = 0.45. 

Based on their location in the CMD, see Figure \ref{Fig3}, we reject two stars from our candidate And XXVII/NW-K1 populations due to their lack of proximity to the isochrone. These are star number 80 on mask A27sf1 and star number 65 on mask 604HaS.

To further refine our selection of candidate And XXVII/NW-K1 stars, we return to their velocities.  To determine which stellar population a given star of velocity, $v_i$ and a velocity uncertainty of $v_{\rm err,i}$ is most likely to belong to, we define a single Gaussian function for each of them of the form:
\begin{equation} 
	\label{eq:3}
	\begin{multlined}
		P_{\rm struc}  = \frac{1}{\sqrt{ 2 \pi( \sigma_{v,\rm struc}^2 + v^2_{{\rm err},i} + \sigma_{\rm sys}^2)}}
		 \times \\  \\
		\shoveleft[1cm] \mathrm{exp}\Bigg[-\frac{1}{2} 
		\bigg( \frac{v_{r,\rm struc} - v_{r,i}} {\sqrt{\sigma_{v,\rm struc}^2 + v^2_{{\rm err},i} + \sigma_{\rm sys}^2)}}
		\bigg)^2 
		\Bigg]
	\end{multlined}
\end{equation}
where: $P$$_{\rm struc}$ is the resulting probability distribution function (pdf); $v$$_{\rm struc}$ {\kms} is the systemic velocity;  $\sigma$$_{v,\rm struc}$ {\kms} is the velocity dispersion and $\sigma_{\rm sys}$ is a systematic uncertainty component of 2.2 {\kms}, determined by \cite{RefWorks:293}, \cite{RefWorks:157}  and \cite{RefWorks:92}\footnote{In calculating the uncertainties for the radial velocity measurements and telluric corrections of stars in their datasets, these authors find that an additional term, $\sigma_{\rm sys}$,  is required to obtain an accurate Gaussian distribution for their data.  Repeated measurements of independent observations determine this  to be 2.2 {\kms} to take into account systematics not included in their Monte Carlo analyses.}, and contemporary with our observations. The likelihood function for membership of And XXVII, based on velocity, is then defined as:
\begin{equation} 
	\label{eq:5}
	\begin{multlined}
		\mathrm{log}[\mathcal{L}(v_r, \sigma_r)] = \sum_{i=1}^{N} \mathrm{log}(\eta_{M31} P_{i,M31} +\\
		 \shoveleft[3cm]  \eta_{MW} P_{i,MW} + \eta_{A27} P_{i,A27})
	\end{multlined}
\end{equation}
\\
where $\eta_{M31}$, $\eta_{MW}$ and $\eta_{A27}$ are the fraction of stars within each stellar population, v$_r$ includes v$_{r \rm A27}$, v$_{r \rm M31}$ and v$_{r \rm MW}$ and $\sigma$$_r$ includes $\sigma$$_{r \rm A27}$, $\sigma$$_{r \rm M31}$ and $\sigma$$_{r \rm MW}$.

We incorporate the above equations, tailored for each stellar population (i.e. MW, M31, And XXVII/NW-K1) into an MCMC analysis, using the {\sc emcee} software algorithm, \cite{RefWorks:571}, \cite{RefWorks:63}. We set the initial values of the systemic velocity to previously published values, with the initial values for the velocity dispersion for And XXVII and the MW based on the spread of velocities in respective stellar populations. For M31 we calculate the initial value for the velocity dispersion in accordance with the the distance, $R$, of the centre of the mask from the centre of M31 (see \citealt {RefWorks:10} and \citealt {RefWorks:235}) given by:
\begin{equation} 
	\label{eq:4}
	\sigma_{v}(R) = \bigg(152 - 0.9 \frac{R}{1\: \rm{kpc}}\bigg)  \: \rm km s^{-1}  kpc^{-1}
\end{equation}
We base the initial values of the fraction parameters $\eta_{M31}$, $\eta_{MW}$ and $\eta_{A27}$ on the distribution of stars plotted on velocity histograms for each mask.

We set broad priors for each stellar populations on the masks i.e. :
\begin{itemize}
	\item systemic velocities are -580 $\le$ $v_{\rm A27}$ /{\kms} $\le$ -450,  -400 $\le$ $v_{\rm M31}$ /{\kms} $\le$ -200 and -150 $\le$ $v_{\rm MW}$ /{\kms} $\le$ -10.
	\item velocity dispersions are 0 $\le$ $\sigma$$_{v \rm A27}$ /{\kms} $\le$ 30, 0 $\le$ $\sigma$$_{v \rm M31}$ /{\kms} $\le$ 100 (except for masks A27sf1 and 603HaS, where the prior is defined as 0 $\le$ $\sigma$$_{v \rm M31}$ /{\kms} $\le$ 200) ; and 0 $\le$ $\sigma$$_{v \rm MW}$ /{\kms} $\le$ 150.
	\item the fraction parameters are 0 $\le$ $\eta$ $\le$ 1 with $\eta_{\rm A27}$ + $\eta_{\rm M31}$ + $\eta_{\rm MW}$  = 1.
\end{itemize}

As part of our analysis, we test the use of broader priors for $\sigma$$_{v \rm A27}$ before adopting those reported above. At values $>$ 30 {\kms} we find the resulting pdfs to be a poor fit for our sparse data and inclusive of M31 halo stars.  We also note an increase in the acceptance fraction which implies a reduction in the quality of the results. So we adopt the range specified, with the upper limit of 30 {\kms} more than twice the expected velocity dispersion for a dwarf galaxy.

We set our Bayesian analysis to run for 100 walkers taking 100000 steps, a burn-in of 1000 and a scale parameter set to the default value of 2.  We use the {\sc emcee} algorithm to fit Gaussians and derive posterior distributions for the systemic velocity, velocity dispersion and the fraction parameters for our stellar populations.  

We find that, for all masks, the MW population can be approximated by a single Gaussian of the above form (Equation \ref{eq:3}) as can the stellar populations for And XXVII and NW-K1. With respect to M31, we find that for masks 7And27, A27sf1, A27sf3 and 604Has there are too few stars to constrain the M31 halo adequately, so we fix the systemic velocity of M31 (at -300 {\kms}), the velocity dispersion (obtained using Equation \ref{eq:4}) and the $\eta_{M31}$ parameter.  For masks 603HaS and A27sf2 we allow the {\sc emcee} algorithm to fit all the data.  The M31 halo is then constrained, or fixed, by a single Gaussian, again of the form shown in Equation \ref{eq:3}. 

We check the Acceptance Fraction of our model to ensure that we have an appropriate number of independent samples to represent the data.  We find, for all models, an Acceptance Fraction $\sim$0.3, which is in the range (0.2 - 0.5) recommended by \cite {RefWorks:382}. 

\begin{figure}
 	\centering
	\includegraphics[width=\columnwidth]{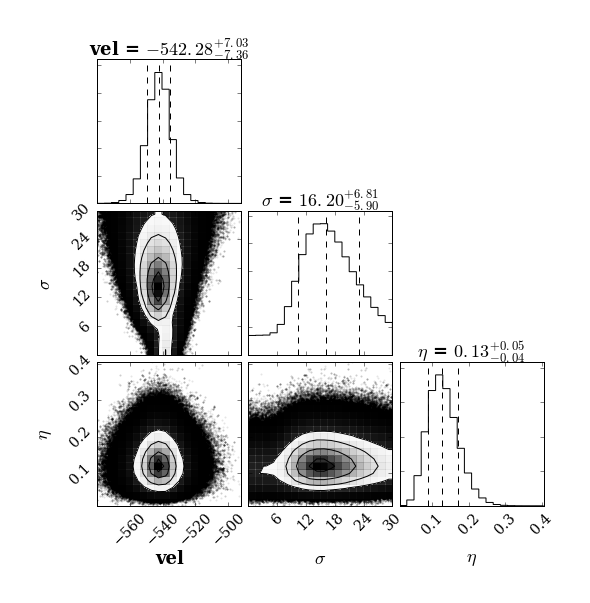}
    	\vspace*{-8.5mm}\caption[2-d marginalised probability distribution functions for mask A27sf1]
	{2-d marginalised probability distribution functions for mask A27sf1 for systemic velocities (vel), velocity distributions ($\sigma$) and percentage of stars in the stellar population ($\eta$) for And XXVII. The dashed lines indicate the median values for each parameter and their 1-$\sigma$ uncertainties. This corner plot is created using code from \cite{RefWorks:511}. }
	\label{Fig4}
\end{figure}

We also check that our MCMC chains are converged using the Integrated Autocorrelation Factor, obtained via the \enquote{acor} function in {\sc emcee}.  For each mask, we determine the precision, p, of the values obtained for each parameter being estimated using: p = $\sqrt{\tau/N}$, where $\tau$ is the autocorrelation time and N is the number of samples. We compare the values obtained ($\sim$0.003 for each mask) with the uncertainties calculated by {\sc emcee} for each parameter.  As the precision is very much smaller than the uncertainties, we are satisfied that the chains are converged.  We then use corner plots, such as the example shown at Figure \ref{Fig4}, to visualise the distribution and covariance of the parameters. 

Having fitted a Gaussian velocity profile for each of the three stellar populations, we derive the probabilities for each star on the masks belonging to a given population using:
\begin{equation} 
	\label{eq:6}
	P_{\rm vel} = \frac{P_{A27}}{P_{M31} +  P_{MW} + P_{A27}}
\end{equation}
\\
with the probability of being a contaminant given by: 
\begin{equation} 
	\label{eq:8}
	P_{\rm contam} = \frac{P_{M31} +  P_{MW}} {P_{M31} +  P_{MW} + P_{A27}}
\end{equation}

% Figure placed here to appear in the right place in the text

Figure \ref{Fig5} shows the velocity distribution histograms overlaid with the pdfs. In each we see a distinct cold peak located in the velocity range of -650 $\le$ v /{\kms} $\le$ -400. Based on these results, we reject one further star, number 16 on mask A27sf1, which has a low probability of association with And XXVII.  

We assess the total probability of stars being members of And XXVII/NW-K1 by combining their probabilities of membership from the CMD and velocity analyses as follows:
\begin{equation} 
	\label{eq:101}
		 P_{\rm A27} = P_{\rm vel}  \times  P_{\rm iso} 
\end{equation}

We identify stars with P$_{\rm A27}$ > 0.6 as members of the stellar populations for And XXVII and NW-K1. We summarise the results of our kinematic analysis in Table \ref{table:7}. 

%--------------------------------------------------------------------------------
\subsubsection{Mask 7And27} \label{Mask 7And27}

\begin{figure}
 	\centering
	\includegraphics[width=\columnwidth]{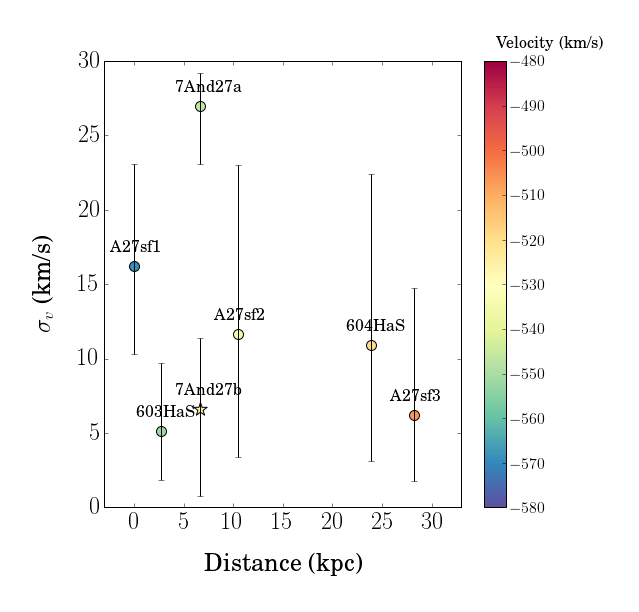}
    \caption[Velocity Dispersions for all masks]
	{Velocity dispersions for all masks with respect to the centre of mask A27sf1 (which lies closest to M31). The values for the velocities and the error bars were obtained from the {\sc emcee} algorithm, the mask locations were obtained using the mean value of all the coordinates for the stars on each respective mask. The datapoint labelled 7And27a represents the velocity dispersion obtained by fitting a single Gaussian to the data, while 7And27b is obtained by isolating the cold peak.}
	\label{Fig40}
\end{figure}

On review of our kinematic analysis of mask 7And27, located across the centre of And  XXVII, we see that the single Gaussian is a poor fit for the data (see Figure \ref{Fig5}). It does not adequately describe the clear cold peak and surrounding hotter components, which could be And XXVII or M31 halo stars.   We also see that while the systemic velocity is in-keeping with the values obtained for the surrounding masks, the velocity dispersion is not (Figure \ref{Fig40}, datapoint 7And27a).

% Placed here to appear in the right place in the text
 \begin{table}
	\centering
	\setlength\extrarowheight{2pt}	
	\caption[Results of the kinematic analysis of And XXVII and NW-K1]
		{Results of the kinematic analysis of And XXVII and NW-K1. For mask 7And27, values are shown for (\textit{a}) the single Gaussian fit to the And XXVII candidate stars and (\textit{b}) the cold peak. Values obtained from the posteriors of the {\sc emcee} algorithm, with uncertainties reported at 68\% confidence limits. }
		\label{table:7}
	\begin{tabular}{lcc} 
		\hline
	 	Mask & v$_r$ & $\sigma_v$ \\
		 & {\kms}& {\kms}\\
		\hline
		And XXVII\\
		7And27$^{(a)}$    &    -526.1  $^{+   10.0  }_{-   11.0  }$    &      27.0  $^{+    2.2  }_{-    3.9  }$\\[1.5ex]
		7And27$^{(b)}$    &    -536.5  $^{+    3.3  }_{-   2.3  }$    &      5.6  $^{+ 6.1  }_{- 4.0  }$\\[1.5ex]
		\hline
		NW-K1\\		
		A27sf1    &    -542.3  $^{+    7.1  }_{-    7.4  }$    &      16.2  $^{+    6.9  }_{-    5.9  }$\\[1.5ex]
		603HaS    &    -530.2  $^{+    2.9  }_{-    3.1  }$    &       5.1  $^{+    4.6  }_{-    3.3  }$\\[1.5ex]
		A27sf2    &    -518.4  $^{+   12.5  }_{-   12.5  }$    &      11.7  $^{+   11.4  }_{-    8.3  }$\\[1.5ex]
		604HaS    &    -507.4  $^{+   10.7  }_{-   10.8  }$    &      10.9  $^{+   11.5  }_{-    7.8  }$\\[1.5ex]
		A27sf3    &    -498.6  $^{+    5.3  }_{-    5.3  }$    &       6.2  $^{+    8.6  }_{-    4.4  }$\\[1.5ex]
		Mean      &    -519.4  $\pm$ 4.0      &      10.0  $\pm$ 4.0 \\[1.5ex]
		\hline
	\end{tabular}
\end{table}

% Placed here to appear in the right place in the text 

\begin{figure}
  	\centering
	\includegraphics[width=0.8\columnwidth]{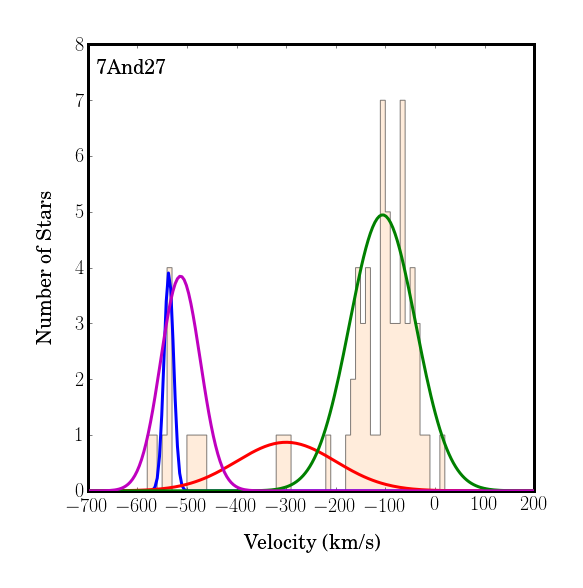}
	\vspace*{-6mm}\caption[Isolating the cold peak on mask 7And27]
	{Velocity histogram overlaid with the membership pdf for each of the three stellar populations - shown in magenta for And XXVII (and equivalent to that shown in Figure \ref{Fig5} for this mask), red for M31 and green for the MW. The blue line indicates the pdf for the And XXVII cold peak. }  
	\label{Fig14}
\end{figure}

We decide to fit two Gaussians to determine whether we could more accurately represent the data and address the above points.   We set broad boundary priors for the hot components and tightly restricted priors to isolate the cold peak. Both Gaussians are given the same initial systemic velocity (-536.0 {\kms}) but different velocity dispersions of 5 {\kms} for the cold and 10 {\kms} for the broad peak. Figure \ref{Fig14} shows the results of this approach. We find the cold peak well constrained with v$_r$ = -536.5$^{+3.3 }_{-2.3 }${\kms} and a velocity dispersion, $\sigma_v$ = 5.6 $^{+6.1 }_{-4.0}$, which is more consistent with the neighbouring fields along the stream (Figure \ref{Fig40}, datapoint 7And27b). 

We note that neither model provides a really good fit for the data so we explore their relative quality using the extended Akaike information criterion, AIC$_c$ (for use with small datasets) and the Bayesian information criterion, BIC, \cite{RefWorks:542}, \cite{RefWorks:527}.  We calculate the values for these estimators using:
\begin{align}
	\label{eq:18}
	AIC_c = -2{\rm log}(\mathcal{L})+ 2K + \frac{2K(K+1)}{n-K-1} 
\end{align}
and:
\begin{align}
	\label{eq:19}
	BIC = -2{\rm ln}(\mathcal{L}) + K
\end{align}
where: $\mathcal{L}$ is the maximum value of the likelihood function for a given model, K is the number of parameters to be estimated and n is the number of data points in the analysis (in our case, the number of stars on the mask).  Neither of these two estimators give any measure of the absolute quality or otherwise of any given model.  Rather they provide insight into the quality of a given model relative to others.  Both estimators encourage simplicity and will naturally favour the model with the fewest parameters to be fitted, so a model with a lower AIC$_c$ or BIC value is deemed to be the ``better'' model.  We find that the AIC$_c$ and the BIC values for both models are of the same order of magnitude, with the single Gaussian model having a marginally ($\sim$ 3\%) lower score for both.   We, therefore, adopt the values from the single Gaussian model for v = -526.1  $^{+10.0}_{-11.0}$ {\kms} and $\sigma_v$ = 27.0$^{+2.2}_{-3.9}${\kms} as the kinematic properties of And XXVII. 

When we compare this to previous works we see that the systemic velocity is $\sim$13 {\kms} lower and $\sigma_v$ $\sim$13 {\kms} higher than those reported by \cite{RefWorks:42}.  This is likely due to our analysis taking fields 603HaS and 7And27 separately rather than combining their data.  Previously thought to lie at the centre of And XXVII, these two fields are $\sim$4 kpc apart on the sky, which is a much larger distance than the r$_h$ for And XXVII (455 pc). As the $\alpha$ and $\delta$ for the centre of mask 7And27 are very close to those for the centre of And XXVII we assign stars in this field to the stellar population for the dSph and assign stars from mask 603HaS to the stellar population for NW-K1. 

The new $\sigma_v$ for And XXVII is also larger than many other dwarf galaxies in the Local Group, with only Canis Major, $\sigma_v$ = 20.0$\pm$3.0 {\kms}, the Large Magellanic Cloud, $\sigma_v$ = 20.2$\pm$0.5 {\kms}, NGC185,  $\sigma_v$ = 24.0$\pm$1.0 {\kms}, the Small Magellanic Cloud, $\sigma_v$ = 27.6$\pm$0.5 {\kms}, NGC205,  $\sigma_v$ = 35.0$\pm$5.0 {\kms} and M32, $\sigma_v$ = 92.0$\pm$5.0 {\kms}, having similar or larger values, \cite{RefWorks:56}. Given that And XXVII has a small r$_h$, we take this large $\sigma_v$ as a possible indicator that And XXVII is being tidally disrupted. 

% Placed here to appear in the right place in the text

While constraining the cold peak we notice four stars (with velocities $\sim$ -500{\kms}, see Figure \ref{Fig5}) that are separate from the main peak and could be affecting our fit of a single Gaussian, causing the large velocity dispersion.  Likely to have been categorised as And XXVII candidates stars due to the sample size and breadth of the priors used for fitting the data, we consider the possibility that these stars could be a substructure within And XXVII. Other possible explanations are:
\begin{itemize}
	\item They are M31 halo stars. With the velocity dispersion, at the distance of this mask, $\sim$99.7{\kms}, these stars are only 2$\sigma$ away from the mean of the M31 halo systemic velocity. However, we see that these stars have a greater than 90\% probability of association with And XXVII, so it is unlikely that they are halo stars.
	\item While these stars may appear to be close to the centre of And XXVII, the imaging data provide only a projected position and they may, in fact, be further away than they seem. 
	\item These stars have been stripped and are no longer in equilibrium with And XXVII. 
\end{itemize}
 
There is no way to know which of the above is correct.  Had the spectra been good enough we might have been able to use [$\alpha$/H] abundances to see if the stars were more consistent with And XXVII or M31.  However, as they have high probabilities of association with And XXVII, there would seem to be some credence to the possibility that they are stripped stars.

%--------------------------------------------------------------------------------
\subsubsection{Velocity Gradients} \label{Velocity Gradients}

Assuming that And XXVII is being tidally disrupted by M31 we look for evidence to support this hypothesis.  In the first instance we explore the possibility of a tangential velocity gradient across the centre of And XXVII. \cite{RefWorks:255} noted that dSphs with such gradients were either undergoing tidal interactions or had an intrinsic rotation. \cite{RefWorks:257} measured a velocity gradient along the major axis of the Hercules dSph and concluded this was indicative of it being pulled apart and transforming into a tidal stream following a close pass of the Milky Way centre.   

Using techniques described by  \cite{RefWorks:257} and \cite{RefWorks:185}, we amend Equation \ref{eq:3} to include a velocity gradient ($\frac{dv}{d\chi}\ $) as shown in \ref{eq:9}.
\begin{equation} 
	\label{eq:9}
	\begin{multlined}
		P_{\rm struc}  = \frac{1}{\sqrt{ 2 \pi( \sigma_{v,\rm struc}^2 + \sigma_{\rm sys}^2)}}
		 \times \\  \\
		\shoveleft[1cm] \mathrm{exp}\Bigg[-\frac{1}{2} 
		\bigg( \frac{\Delta v_{r,i}} {\sqrt{\sigma_{v,\rm struc}^2 + + \sigma_{\rm sys}^2)}}
		\bigg)^2 
		\Bigg]
	\end{multlined}
\end{equation}
where $\Delta v_{r,i}$ ({\kms}) is the velocity difference between the \textit{i}$^{th}$ star and a velocity gradient, $\frac{dv}{d\chi}\ $ ({\kms}arcmins$^{-1}$), acting along the angular distance of the star along an axis, $y_i$ (arcmins), with a position angle, $\theta$ (radians, measured from North to East) and is given by: 
\begin{equation} 
	\label{eq:10}
	\Delta v_{r,i} = v_{r,i}  -  \frac{dv}{d \chi} y_i +  \langle v_{r} \rangle
\end{equation}

We determine $y_i$ using the right ascension and declination of the \textit{i}$^{th}$ star ($\alpha_i$, $\delta_i$) and the centre of And XXVII ($\alpha_0$, $\delta_0$) using:
\begin{align}
	\label{eq:11}
	y_i = X_i \rm sin(\theta) + \mathit {Y_i} \rm cos(\theta)   
\end{align}
where:
\begin{align}
	\label{eq:12}
	X_i  = (\alpha_i - \alpha_0)\rm cos(\delta_0)\qquad   \text{and} \qquad
	\mathit {Y_i} = \delta_i - \delta_0
\end{align}

We define flat priors for these new terms, i.e. -150 $\le$ $\frac{dv}{d\chi} (${\kms}arcmins$^{-1}$) $\le$150 and 0 $\le$ $\theta$ (radians) $\le$ $\pi$ and use the priors from our single Gaussian analysis for the other parameters (\textit v, $\sigma_v$ and $\eta$) before running the {\sc emcee} algorithm on the confirmed stellar populations on masks 7And27, 603HaS, A27sf1 and A27sf3 as they cover the length of the stream and each includes sufficient And XXVII/NW-K1 stars to deliver meaningful results. 

\begin{figure}
  	\centering
	\includegraphics[width=\columnwidth]{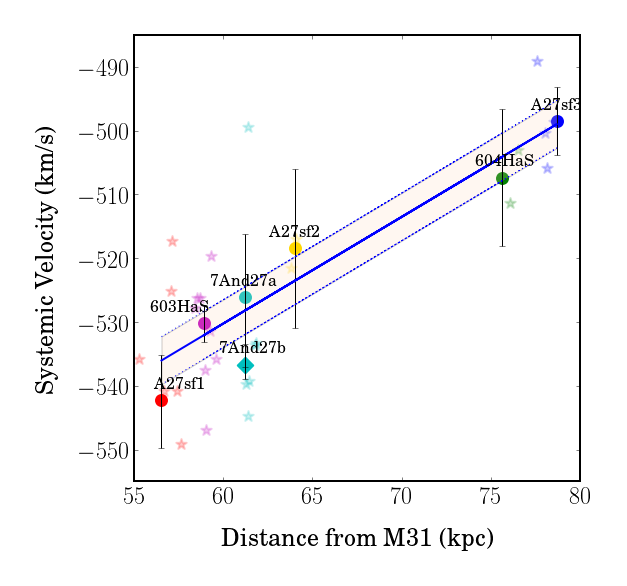}
    	\vspace*{-9mm}\caption[Velocity gradient across stars on all masks with respect to the centre of mask A27sf1]
	{Velocity gradient across all masks with respect to the centre of M31. The values for the velocities and the uncertainties (at 68\% confidence limits) were obtained from the {\sc emcee} algorithm. The mask locations were obtained using the mean value of all the $\alpha$s and $\delta$s for the stars on each respective mask. The blue line is the best fit line and has a gradient of -1.7$\pm$0.3 {\kms} kpc$^{-1}$.  The shaded area, bounded by dotted lines, indicates the standard deviation ($\pm$3.7 {\kms}) about the best fit line. The star shaped icons represent the And XXVII/NW-K1 confirmed stellar population for each mask, the colour coding denotes the mask on which they were observed.}
	\label{Fig7}
\end{figure}

Our findings show marginal tangential gradients in all four masks but none are significant or have any effect on the systemic velocities and dispersions recorded in Table \ref{table:7}. This could, again, be due to the sample size on each mask or could indicate that we are seeing stars, moving at high velocities along a straight line path, that are the debris of a completely unbound system.  If this latter is the case, this could also explain the different velocity dispersions we obtain from our analysis of mask 7And27.  Simulations by \cite{RefWorks:489} find that data taken from particular lines of sight can be contaminated by unbound stars in tidal tails, 

Next we look to see if there is a velocity gradient across And XXVII and NW-K1. In Figure \ref{Fig7} we present the systemic velocity as a function of the  distance of each mask from the centre of M31. Based on line of sight data, this plot shows a distinctive velocity gradient (-1.7$\pm$0.3 {\kms} kpc$^{-1}$) that becomes increasingly negative in the direction of M31 - indicative of an infall trajectory. The small scatter of the mask velocities around the best fit line implies that the stars belong to a dynamically cold system.

We compare these findings with those of \cite{RefWorks:72}, \cite{RefWorks:232} and \cite{RefWorks:236} on Globular Clusters (GC) projected onto NW-K2. \cite{RefWorks:236} find a velocity gradient of -1.0$\pm$0.1{\kms} kpc$^{-1}$ across the GCs, that becomes increasingly negative in the direction of M31.  

With both NW-K1 and NW-K2 appearing to have trajectories infalling towards M31 we hypothesise that they cannot be part of the same structure. However, we note that as we only have line of sight velocities, the gradients could be due to the shape of the stream and there could be stronger, undetectable, velocity components in directions compatible with a single structure.  Since \cite{RefWorks:405} find both sections of the NW stream lie behind M31 for them both to have infall trajectories towards M31 is difficult to reconcile with a single stellar structure and it is more probable that they are different streams.

%--------------------------------------------------------------------------------
\subsection{Metallicities}\label{Metallicities}

We measure the spectroscopic metallicities of the stars in And XXVII /NW-K1 using the CaT lines between 8400\AA-8700\AA.  As the S/N  > 3 for only 12 stars, we conclude that using the individual spectra to determine metallicities would not deliver robust results.

In low S/N spectra, the CaT lines are hard to distinguish from the noise and are, therefore, unreliable. As we see only a small spread of metallicities on the CMD, we choose to stack the spectra (following the approach of \citealt{RefWorks:444}, \citealt{RefWorks:160, RefWorks:175} and \citealt{RefWorks:52, RefWorks:171}) as combining them will likely give a good estimate of the average. We acknowledge that the approach may deliver a more metal-rich result than expected while recognising that the technique provides a good indication of mean values. 

\begin{figure}
  	\centering
	\includegraphics[width=\columnwidth]{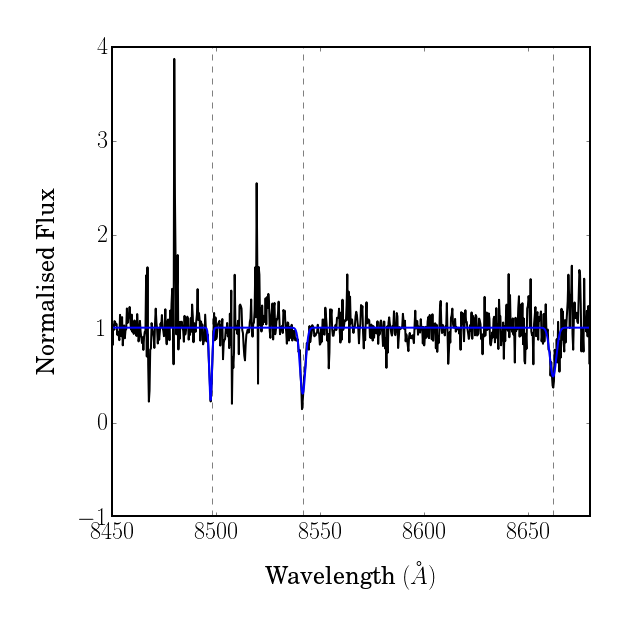}
    	\vspace*{-9mm}\caption[Coadded, weighted, spectrum from mask 603HaS]
	{Co-added spectrum from mask 603HaS overlaid with the best fit curve (solid blue line). This is representative of the results of the spectral analyses conducted for all masks. The dotted lines indicate the position of the CaT.}
	\label{Fig9}
\end{figure}

We prepare the individual spectra, following the method described by  \cite{RefWorks:42}, by correcting for the velocity of the individual star, smoothing the spectrum and normalising the data using a median filter.  We weight each spectrum by its S/N and interpolate the spectra to a consistent framework before co-adding the fluxes. We simultaneously fit the continuum and the CaT lines to the co-added spectrum (see example shown at Figure \ref{Fig9}) to obtain the equivalent widths.

The relationship between equivalent width and [Fe/H] of an object is well established (\citealt{RefWorks:612}, \citealt{RefWorks:610},  \citealt{RefWorks:272}) and has been calibrated for use with all three, or subsets thereof, CaT lines. As not all of our co-added spectra have three well defined CaT lines (i.e. our first line is often contaminated by sky-lines) and as the CMD indicates our stars are metal-poor, we follow the approach described by \cite{RefWorks:272} and substitute the equivalent widths obtained above into:

\begin{equation} 
	\label{eq:21}
	\begin{multlined}
		[\mathrm{Fe/H]} = a + bM + cEW_{(2+3)} + dEW_{(2+3)}^{-1.5} + eEW_{(2+3)}M   
	\end{multlined}
\end{equation}
\\
where: $a$, $b$, $c$, $d$ and $e$ are taken from the calibration to the Johnson-Cousins \textit{M$_I$} values and equal to -2.78, 0.193, 0.442, -0.834 and 0.0017 respectively; and EW$_2$ and EW$_3$ are the the equivalent widths for the CaT lines at 8542\AA{} and  8662\AA {} respectively. EW$_{(2+3)}$ = EW$_2$ and EW$_3$. $M$ is the absolute magnitude of the star given by: 
\begin{equation} 
	\label{eq:22}
	M = i - 5 \times \mathrm{log}{_{10}}(\mathrm{D}{_{\odot}}) + 5
\end{equation}	
where: $i$ is the i-magnitude of the star and $D{_{\odot}}$ is the heliocentric distance for the star, which we assume, in all cases, to be the heliocentric distance for And XXVII.  Uncertainties on the metallicity are determined using the uncertainties on the equivalent widths, obtained from the covariance matrix produced by the fitting process, combined in quadrature.  Other intrinsic uncertainties in the approach taken are discussed in \cite{RefWorks:272}.

% Placed here to appear in the right place in the text

We present our results in Table \ref{table:2}.   We find the metallicity for And XXVII to be -2.1$^{+0.4}_{-0.5}$, while NW-K1 has a mean metallicity of -1.8$\pm$0.4.  Our values are consistent with previous findings of: [Fe/H]$_{\rm phot}$ = -1.7$\pm$0.2, \cite{RefWorks:18}; [Fe/H]$_{\rm spec}$ = -2.1$\pm$0.5, \cite{RefWorks:42}; the metallicity of And XXVII's RR Lyrae stars = -1.62$\pm$0.23, \cite{RefWorks:474}, and with the results from the isochrone analysis.

\begin{figure}
  	\centering
	\includegraphics[width=\columnwidth]{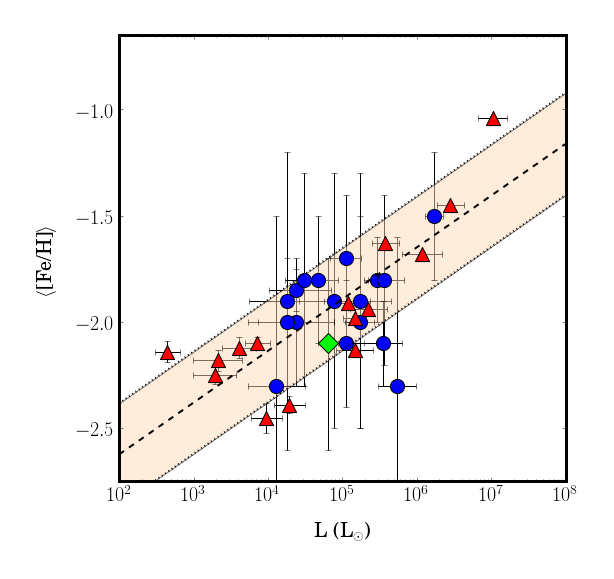}
    	\vspace*{-9mm}\caption[Spectroscopic metallicities as a function of luminosity at the half light radius]
	{Spectroscopic metallicities as a function of luminosity at the half light radius for Local Group dwarf galaxies with those from M31 shown as blue circles, those from MW shown as red triangles and And XXVII/NW-K1 shown as a green diamond. The dashed line indicates the best fit line, following the approach by \cite{RefWorks:362}.  The shaded area, bounded by dotted lines, indicates the 1$\sigma$ deviation. Data sources: \cite{RefWorks:42} and \cite{RefWorks:362} and \cite{RefWorks:512}.}
	\label{Fig13}
\end{figure}

We plot the metallicity of And XXVII as a function of luminosity (Figure \ref{Fig13}). \cite{RefWorks:362} established that the metallicities of MW dSphs decrease with decreasing luminosity in a relationship given by: 
\begin{equation} 
	\label{eq:20}
	\langle\mathrm{[Fe/H]\rangle_{dSph} =  (-1.69\pm0.06)+(0.29\pm0.04)log}\Bigg( \frac{L_v}{10^6L_\odot}\Bigg)
\end{equation}
where: $\langle$[Fe/H]$\rangle$$_{\rm dSph}$ is the average weighted mean metallicity of the galaxy and L$_v$ is its corresponding luminosity. \cite{RefWorks:42} showed that the M31 dSphs also conform to this relationship.  Our plot shows that And XXVII's metallicity lies within 1-$\sigma$ of the best fit line and is consistent with Local Group dwarf galaxies of similar luminosities.

% Table placed here to be in the right place in the paper
\begin{table}
	\centering
	\setlength\extrarowheight{2pt}	
	\caption[Metallicities ]
	{Metallicities obtained from co-added spectra weighted by $S/N$ for stars on each mask.}		
	\label{table:2}
	\begin{tabular}{lc} 
		\hline
		{Mask} & [Fe/H]$_{\rm spec}$  \\  
		\hline
		And XXVII &  \\
		7And27     &   -2.1$^{+0.4}_{-0.5}$  \\ [1.5ex] 
		\hline
		NW-K1     &  \\	
		A27sf1     &  -1.7$^{+0.3}_{-0.4} $  \\ [1.5ex]       
		603HaS   &  -1.5$\pm$ 0.4              \\ [1.5ex]              
		A27sf2     &  -1.6$^{+1.5}_{-4.0}  $  \\ [1.5ex]      
		604HaS   &  -2.5$^{+0.6}_{-1.0}  $  \\ [1.5ex]      
		A27sf3     &  -1.8$\pm$0.5                \\ [1.5ex]             		           
                \hline
                 NW-K1 mean  &  -1.4$\pm$0.1   \\ [1.5ex]                  
		\hline
	\end{tabular}
\end{table}

% Placed here to appear in the right place in the text
\begin{table*}
	\centering	
	\setlength\extrarowheight{1pt}	
	\vspace*{-5mm}\caption[Properties of the And XXVII and NW stream stars]
	{Table showing properties of And XXVII and NW-K1 candidate stars. A similar catalogue of all our observed stars is provided electronically with this article. The columns include: (1) Star number; (2) Right Ascension in J2000; (3) Declination in J2000; (4) \textit{i}-band magnitude; (5) \textit{g}-band magnitude; (6) \textit{S/N} (\AA$^{-1}$); (7) line of sight heliocentric velocity, \textit v ({\kms}); (8) Probability of membership of And XXVII and NW-K1 as defined in Equation \ref{eq:101}. }
	\label{table:3} 
	\begin{tabular}{lccccclr} 
		\hline
			Mask/star  & $\alpha$ & $\delta$ & \textit{i} & \textit{g} & \textit{S/N} &    \multicolumn{1}{c}{\textit{v$_r$}}   & \textit{P$_{A27}$}   \\  
		\hline
		And XXVII\\
				7And27 & & & & & &  &  \\ 
 5  & 00:37:18.84 & +45:23:19.3 &  22.1  &  23.3  &  2.5   & -463.2 $\pm$ 4.8  &    0.65 \\
 6  & 00:37:19.75 & +45:23:51.8 &  21.4  &  22.8  &  4.7   & -539.6 $\pm$ 4.0  &   0.98 \\
 7  & 00:37:19.82 & +45:24:17.8 &  21.6  &  23.0  &  3.9   & -476.1 $\pm$ 6.4  &   0.83 \\
11  & 00:37:19.24 & +45:21:36.4 &  21.8  &  23.1  &  3.1   & -563.1 $\pm$ 4.1  &   0.97 \\
19  & 00:37:33.79 & +45:25:18.9 &  22.4  &  23.6  &  2.2   & -539.2 $\pm$ 9.6  &   0.99 \\
20  & 00:37:41.84 & +45:25:27.9 &  22.0  &  23.5  &  2.7   & -544.7 $\pm$ 5.2  &   0.94 \\
31  & 00:37:36.90 & +45:27:06.8  &  21.8  &  23.3  &  3.5   & -533.3 $\pm$ 5.3  &    0.96 \\
32  & 00:37:43.68 & +45:27:11.3 &  21.3  &  23.0  &  5.3   & -533.5 $\pm$ 3.2  &   0.97 \\
46  & 00:37:7.86 & +45:22:50.5 &  22.6  &  23.8  &  2.0   & -579.1 $\pm$ 15.6 &   0.95 \\
54  & 00:37:21.21 & +45:24:25.2 &  23.2  &  24.1  &  1.5   & -499.5 $\pm$ 11.4 &   0.95 \\
55  & 00:37:19.59 & +45:24:37.5 &  23.2  &  24.2  &  0.8   & -480.4 $\pm$ 4.0  &   0.91 \\
		\hline
NW-K1\\
		A27sf1 & & & & & & &   \\ 
11  & 00:39:28.22 & +45:09:7.8  &  21.3  &  23.0  &  5.3   & -540.9 $\pm$ 3.2  &   0.87 \\
23  & 00:40:12.21 & +45:04:21.1 &  22.1  &  23.4  &  2.2   & -535.8 $\pm$ 7.6  &   0.97 \\
31  & 00:39:37.65 & +45:8 :52.8 &  22.8  &  24.0  &  1.3   & -560.0 $\pm$ 16.5 &    0.98 \\
32  & 00:39:24.05 & +45:10:26.6 &  22.9  &  23.9  &  1.4   & -571.2 $\pm$ 5.6  &    0.98 \\
33  & 00:39:18.52 & +45:10:29.4 &  22.8  &  24.0  &  1.8   & -525.1 $\pm$ 12.1 &  0.92 \\
35  & 00:39:30.70 & +45:11:07.3  &  22.7  &  24.0  &  1.6   & -517.3 $\pm$ 11.0 &   0.9  \\
37  & 00:39:22.19 & +45:11:56.9 &  22.6  &  24.0  &  1.6   & -540.9 $\pm$ 9.6  &   0.83 \\
38  & 00:39:25.79 & +45:12:53.9 &  22.6  &  23.9  &  1.4   & -549.1 $\pm$ 16.7 &   0.93 \\

\\		
		603HaS & & & & & & &   \\ 
10  & 00:39:8.53 & +45:15:46.8 &  21.2  &  23.0  &  4.6   & -527.9 $\pm$ 5.0  &   0.86 \\
15  & 00:39:5.93 & +45:16:55.3 &  22.1  &  23.4  &  2.2   & -526.2 $\pm$ 3.1  &   0.99 \\
20  & 00:39:25.73 & +45:19:55.0 &  22.4  &  23.6  &  1.5   & -531.5 $\pm$ 4.0  &   1.0  \\
32  & 00:38:30.15 & +45:18:18.1 &  21.7  &  23.1  &  3.0   & -519.6 $\pm$ 6.0  &   0.99 \\
33  & 00:38:46.02 & +45:17:28.4 &  21.2  &  23.2  &  4.8   & -537.5 $\pm$ 4.0  &   0.6  \\
35  & 00:38:38.75 & +45:17:33.4 &  21.7  &  23.2  &  3.4   & -546.9 $\pm$ 8.1  &   0.98 \\
38  & 00:38:44.39 & +45:15:36.2 &  21.7  &  23.3  &  3.6   & -526.2 $\pm$ 5.2  &   0.91 \\
46  & 00:38:32.03 & +45:19:34.1 &  22.3  &  23.7  &  1.7   & -535.8 $\pm$ 17.3 &   0.9  \\
		\\
		A27sf2 & & & & & & &   \\ 
33  & 00:36:4.85 & +45:31:17.8 &  22.6  &  24.0  &  1.5   & -521.6 $\pm$ 15.5 &   0.9  \\
42  & 00:36:42.08 & +45:34:11.6 &  22.5  &  23.7  &  1.1   & -517.0 $\pm$ 8.3  &   1.0  \\
 		\\
		604HaS & & & & & & &   \\  
16  & 00:31:44.15 & +46:11:9.8  &  21.6  &  23.0  &  2.7   & -503.1 $\pm$ 9.1  &    0.99 \\
 		\\ 
		A27sf3 & & & & & & &   \\ 
 7  & 00:30:33.8 & +46:10:30.1 &  21.6  &  23.1  &  8.7   & -489.1 $\pm$ 7.9  &    0.98 \\
31  & 00:30:33.67 & +46:13:04.5  &  22.2  &  23.6  &  2.4   & -505.8 $\pm$ 7.3  &   0.93 \\
32  & 00:30:45.72 & +46:13:25.7 &  22.3  &  23.7  &  2.0   & -500.4 $\pm$ 11.3 &   0.96 \\
35  & 00:30:32.56 & +46:14:45.8 &  22.6  &  23.7  &  1.6   & -498.7 $\pm$ 3.8  &    0.97 \\
		\hline
		\vspace*{-6.0mm}
	\end{tabular}
\end{table*} 

%%%%%%%%%%%%%%%%%%%%%%%%%%%%%%%%%%%%%%%%%%%%%%%%%%%%%%%%%%%%%%%%%%%%%%%%%%
%%%%%%%%%%%%%%%%%%%%%%%%%%%%%%%%%%%%%%%%%%%%%%%%%%%%%%%%%%%%%%%%%%%%%%%%%%
\section{Conclusions} \label{Conclusions} 
We present our kinematic and spectroscopic analysis of 38 red giant branch stars from 7 fields, including 4 new ones, spanning And XXVII and NW-K1.  We have confirmed secure members of the stellar populations belonging to both features defined by strong ($\ge$ 1$\sigma$) association with our fiducial isochrone and with the systemic velocity for And XXVII (see Table \ref{table:3}).  Our results lead us to conclude:  
\begin{itemize}
	\item \textit{The heliocentric distance for And XXVII is 827 $\pm$ 47 kpc}.  Determined by \cite{RefWorks:18} this was later revised by \cite{RefWorks:19} to 1255 $^{+42}_{-474}$ kpc using their Tip of the RGB approach.  However, their work also shows the distance posterior distribution for And XXVII has a pronounced second peak, indicative of an heliocentric distance $\sim$800 kpc.  Given this value is consistent with the findings of \citeauthor{RefWorks:18} and \cite{RefWorks:474} and that our confirmed member stars for And XXVII are probably brighter than assumed by \citeauthor{RefWorks:19} (as can be seen in the CMD plot for mask 7And27 at Figure \ref{Fig3}, where the dashed isochrone has been distance corrected to 1255 kpc) we believe the original heliocentric distance to be the most likely.
	\item \textit{And XXVII is likely to be in the process of being tidally stripped by M31}. We measure $\sigma_v$ for And XXVII to be $\sim$13 {kms} higher than previously reported.  This new $\sigma_v$ is inconsistent with neighbouring NW-K1 fields and with many other dwarf galaxies in the Local Group. However, it is consistent with expectations of a dwarf galaxy in the throes of a tidal disruption event. 
	\item Given the possibility that And XXVII is tidally disrupting and is not in virial equilibrium, we cannot constrain its mass. However, our kinematic results, especially for NW-K1, may be useful for mapping the orbit of And XXVII and determining its pre-infall halo mass (Preston et al, in prep).	
	\item \textit{And XXVII is a plausible candidate to be the progenitor of NW-K1}: We see consistent properties in the stellar populations for And XXVII and NW-K1 both kinematically, with systemic velocities of -542.3$^{+7.1}_{-7.4}$ $\le$ v /{\kms} $\le$ -498.6 $\pm$ 5.3 and spectroscopically with -1.5 $\le$  [Fe/H] $\le$ -2.5.  In addition we find a velocity gradient consistent with an infall trajectory towards M31 across  And XXVII and NW-K1. Taken together, this indicates that And XXVII and NW-K1 are, potentially, elements of a single feature of which And XXVII may very possibly be the progenitor.
	\item \textit{The NW Stream may not be not a single structure}:  The velocity gradients from both NW-K1 and NW-K2 are indicative of infall trajectories towards M31. As both streams may lie behind M31, this is difficult to reconcile with a single trajectory looping around M31. Therefore, we conclude that it more likely that they are different streams, with NW-K1 associated with And XXVII and NW-K2 not. 
\end{itemize}

%%%%%%%%%%%%%%%%%%%%%%%%%%%%%%%%%%%%%%%%%%%%%%%%%%%%%%%%%%%%%%%%%%%%%%%%%%%
%%%%%%%%%%%%%%%%%%%%%%%%%%%%%%%%%%%%%%%%%%%%%%%%%%%%%%%%%%%%%%%%%%%%%%%%%%%
\section{Acknowledgements}
JP wishes to thank Marla Geha for her useful discussions and advice on this work.  We also thank the anonymous referee who's insightful questions and comments enhanced this paper. 

This work used the community-developed software packages: Matplotlib (\citealt{RefWorks:616}), NumPy (\citealt{RefWorks:615}) and Astropy (\citealt{RefWorks:613, RefWorks:614}).

Most of the data presented herein were obtained at the W.M. Keck Observatory, which is operated as a scientific partnership among the California Institute of Technology, the University of California and the National Aeronautics and Space Administration. The Observatory was made possible by the generous financial support of the W.M. Keck Foundation. 

Data were also used from observations obtained with MegaPrime/MegaCam, a joint project of CFHT and CEA/DAPNIA, at the Canada-France-Hawaii Telescope which is operated by the National Research Council of Canada, the Institut National des Sciences de l'Univers of the Centre National de la Recherche Scientifique of France, and the University of Hawaii.	

The authors wish to recognise and acknowledge the very significant cultural role and reverence that the summit of Mauna Kea has always had within the indigenous Hawaiian community. 
\\
\\
%%%%%%%%%%%%%%%%%%%%%%%%%%%%%%%%%%%%%%%%%%%%%%%%%%%%%%%%%%%%%%%%%%%%%%
%%%%%%%%%%%%%%%%%%%%%%%%%%%%%%%%%%%%%%%%%%%%%%%%%%%%%%%%%%%%%%%%%%%%%%	
%------------------------------------------------------------------- Bibliography  ------------------------------------------------------------------------------------
\bibliography{Biblog27}
\bibliographystyle{mnras}

\end{document}